\documentclass[10pt]{article}
\usepackage{amssymb}
\usepackage{amsmath}
\usepackage{wasysym}
\usepackage{graphicx}
\begin{document}
	
	\title{Born's rule and permutation invariance}
	
	\author{C. Dedes\thanks{{c\_dedes@yahoo.com}}\\
		Operam Education Group  \\
		3 Morston Claycliffe Office Park \\
		Whaley Road, Barnsley, S75 1HQ\\
		United Kingdom \\}
	
	\maketitle
	
	\begin{abstract}
		It is shown that the probability density satisfies a hyperbolic equation of motion with the unique characteristic that in its many-particle form it contains derivatives acting at spatially remote regions. Based on this feature we explore inter-particle correlations and the relation between the quantum equilibrium condition and the permutation invariance of the probability density. Some remarks with respect to the quantum to classical transition are also presented.
		
	\end{abstract}

	\section{Introduction}
	
	A major difference between quantum mechanics and classical physics is the violation of the spatial separability principle \cite{Cushing}. For some physicists like Einstein this even undermined the possibility of doing science and establishing physical laws. It must be noted nevertheless that this distinct quality of quantum physics is reflected only on the entangled form of the total wavefunction of a many particle system. On the other hand the dynamical equation that governs it exhibits exactly the same kind of system separability as in the classical case. Indeed, from a mereological point of view the Hamiltonian of a many-particle quantum system is exactly as a classical one. In the present article without modifying the fundamental equation of motion for the wavefunction we deduce a wave equation for the probability density which will make apparent the holistic aspect of quantum theory. In the next section we present the single particle formalism and examine the conditions that may modify the quantum mechanical predictions. This is related to the validity of the quantum equilibrium condition $\rho=|\Psi|^{2}$. In section 3 we examine the two-particle formulation and the connection between the quantum equilibrium condition and the permutation invariance principle and we conclude with the final remarks.

	\section{Single-particle formalism}
	
	 It has been noted by Dyson \cite{Dyson} that in close analogy to Maxwell's theory, the structure of quantum theory is two-fold, with two distinct layers of description. The more fundamental substrate constitutes the first layer which contains abstract wavefunctions, whereas the second one is more concrete and contains directly observable quantities like probabilities and intensities. In the present section we derive a hyperbolic equation that governs the probability density and also includes spatial and time derivatives of the velocity field.

	Considering then the single-particle one-dimensional Schr\"{o}dinger equation
	
	\begin{equation}
	i\hbar\partial _{t}\Psi = 
	\left(-\frac{\hbar ^{2}}{2m}\partial ^{2}_{x}+V(x,t)\right)\Psi , 
	\end{equation}
	
	\noindent
	where $V(x,t)$ the time-dependent external potential and $\Psi$ a wavefunction in the coordinate representation, we will perform a Madelung transformation
	
	\begin{equation}
	\Psi=\sqrt{\rho}e^{iS/\hbar},
	\end{equation}
	
	\noindent
	where $S/\hbar$ is the phase of the wavefunction and the velocity field is given by the guiding equation
	
	\begin{equation}
	v=\frac{1}{m}\partial_{x} S.
	\end{equation}

	\noindent
	After separating real and imaginary parts two formulas are obtained. By equating the imaginary parts we arrive at local continuity equation for the probability density

	\begin{equation}
	\partial_{t} \rho+\partial_{x} (\rho v)=0. 
	\end{equation}
	
	\noindent
	Consequently the Hamilton-Jacobi equation is obtained from (1) through (2) by equating its real parts, 
	
	\begin{equation}
	- \partial_{t} S-\frac{1}{2m}\left(\partial _{x} S\right)^{2}=V+Q.
	\end{equation}
	
	\noindent
	In the above expression
	
	\begin{equation}
	Q=-\frac{\hbar ^{2}}{2m}\frac{\partial ^{2}_{x}\sqrt{\rho}}{ \sqrt{\rho}}=-\frac{\hbar^{2}}{4m\rho}\left[\partial ^{2}_{x}\rho-\frac{(\partial _{x} \rho)^{2}}{2\rho}\right]  
	\end{equation}

	\noindent
	is the Bohmian quantum potential. Taking the divergence gives the second hydrodynamic equation 
	
	\begin{equation}
	\partial _{t} v+v\partial v_{x}=-\frac{1}{m}\partial _{x}(V+Q),
	\end{equation}
	
	\noindent
	from which it follows \cite{Wyatt, Bohm, Holland, Cushing2}

	\begin{equation}
	\partial _{t}( \rho v)=-\frac{1}{m}\partial _{x} \Pi-\frac{\rho}{m} \partial_{x} V,
	\end{equation}

	\noindent
	where the probability quantum stress tensor is written as

	\begin{equation}
	\Pi=\rho v^{2}+\frac{\hbar^{2}}{4m^{2}}\left(\frac{\partial _{x}ln \rho}{\rho}-\partial ^{2}_{x}\rho\right).
	\end{equation}

	\noindent
	Differentiating (4) over time and (8) over the spatial dimension we finally this wave equation for the probability density

	\begin{equation}
	\partial ^{2}_{t}\rho =\partial ^{2}_{x}\Pi+\partial _{x}\left(\rho\partial_{x} V\right).
	\end{equation}  
	
	\noindent
	An equation of this general form was derived by Lighthill in the study of aeroacoustics \cite{aeroacoustics} and by substituting $\rho=e^{w}$ in the above expression we can ensure its positivity. We note that (10) is nonlinear even though we started from a linear equation for $\Psi$ and ofcourse the local continuity equation for the probability density is linear too. It must be also remembered that at least in the Bohmian conceptual framework the probability density has only a contingent and not a necessary relation to $|\Psi|^{2}$. Even if it happens that these two quantities coincide numerically they refer to two distinct realities in the quantum mechanical formalism. As it is evident from the polar substitution (10) satisfies the quantum equilibrium condition. In a way we have used Born's rule as a heuristic tool in order to deduce a second order equation in time. At the same time, (10) admits also solutions that violate this condition, so $\rho \neq |\Psi|^{2}$. Our task is to examine if we should accept or exclude such a family of solutions. Definitely, the ultimate \textit{desideratum} of a physical theory is agreement with observation and experiment and this clearly dictates that the quantum equilibrium condition must be fulfilled but as we will see in the following section there are additional fundamental considerations for that related to permutation symmetry. So, we will assume that the quantum equilibrium condition (or hypothesis) always holds which also means that the Hamilton-Jacobi equation is valid by setting $\rho =|\Psi|^{2}$ and the local velocity (3). So after repeating the above steps having substituted (5) for the sum of the classical and quantum potentials we will find

	\begin{equation}
	\partial ^{2}_{t} |\Psi|^{2}=\partial ^{2}_{x}\left(|\Psi|^{2}v^{2}\right)-\partial_{ x}\left[|\Psi|^{2}(\partial _{t}+v\partial _{x})v\right],
	\end{equation}

	\noindent
	where we have included the material derivative. The above is a linear equation with respect to $|\Psi|^{2}$, in contradistinction to the non-linear (10), satisfies by construction the quantum equilibrium condition, the guiding equation for the velocity field and the quantum Hamilton-Jacobi equation. As with the continuity equation (3) it does not depend on the applied potential but only on the velocity field and its time derivative and gradient. It relates the second derivative of the probability density with the velocity of the probability fluid without any reference to the classical or quantum potential and is subject to appropriate initial conditions for $|\Psi(x,0)|^{2}$ and $\partial _{t}|\Psi(x,0)|^{2}$. This is the formula we will extend and investigate the consequences of the two-particle case in the next section. As a side note we could observe that in three dimensions the second spatial derivative term includes the dyad of the velocity and since $\frac{1}{2}\nabla \mathbf{v}^{2}=(\mathbf{v}\cdot \nabla)\mathbf{v}+\mathbf{v}\times(\nabla \times \mathbf{v})$ it follows that the probability density displays dependence on the velocity field even if the latter is irrotational and coupling to a vector potential leads immediately to the Aharonov-Bohm effect. We should also note that according to (11) the velocity field is uniquely defined so it is not possible to add a divergence-less velocity field term to the guiding equation as in \cite{Ghirardi}.

	\section{Permutation invariance for a bi-particle quantum compound}

	The two-particle case is more interesting from a conceptual point of view. We consider a system of two interacting spinless particles with equal masses in one dimension. Following the derivation in \cite{Wyatt} for the single particle case, we derive the probability density local continuity equation  
	
	\begin{equation}
	\partial _{t}|\Psi _{12}|^{2}+\partial_{ x_{1}}(|\Psi _{12}|^{2} v_{1})+\partial_{ x_{2}}(|\Psi _{12}|^{2} v_{2})=0
	\end{equation}
	
	\noindent
	where $\Psi =\Psi _{12}$ the two-body probability amplitude. The corresponding two-body Hamilton-Jacobi equation

	\begin{equation}
	-\partial_{t} S_{12}-\frac{1}{2m_{1}}\left(\partial _{x_{1}} S_{12}\right)^{2}-\frac{1}{2m_{2}}\left(\partial _{x_{2}} S_{12}\right)^{2}=V+Q.
	\end{equation}
	
	\noindent
	where the $V$, $Q$ the two-particle classical and quantum potentials. The two corresponding probability field velocities are expressed as 
	
	\begin{equation}
	v_{i}=\frac{1}{m_{i}}\partial _{x_{i}}S_{12}, i=1,2,
	\end{equation}
	
	\noindent
	and we also note that 
	
	\begin{equation}
	\frac{1}{m_{2}} \partial_{x_{2}} v_{1} =\frac{1}{m_{1}} \partial_{x_{1}} v_{2}.
	\end{equation}
	
	\noindent
	Taking the two spatial derivatives of (13) yields
	
	\begin{equation}
		\partial _{t} v_{1}= -v_{1}\partial_{x_{1}} v_{1}-v_{2}\partial_{x_{1}} v_{2}-\frac{1}{m_{1}}\partial_{x_{1}}(V+Q),
	\end{equation}

	\begin{equation}
	\partial _{t} v_{2}= -v_{2}\partial_{x_{2}} v_{2}-v_{1}\partial_{x_{2}} v_{1}-\frac{1}{m_{1}}\partial_{x_{2}}(V+Q).
	\end{equation}

	\noindent
	Employing the above and the continuity equation we obtain these Navier-Stokes equations for the momentum fields
	
	\begin{align}
	    	\partial_{t} (|\Psi |^{2}v_{1} )=& -\partial_{x_{1}} (|\Psi|^{2} v^{2}_{1})-|\Psi|^{2}v_{2}\partial _{x_{2}}v_{1} \nonumber\\
	    	& -v_{1}\partial_{x_{2}}(|\Psi |^{2}v_{2})-\frac{|\Psi |^{2}}{m_{1}} \partial_{x_{1}} (V+Q),
	\end{align}

	\begin{align}
	    	\partial_{t} (|\Psi |^{2}v_{2} )=& -\partial_{x_{2}} (|\Psi|^{2} v^{2}_{2})-|\Psi|^{2}v_{1}\partial _{x_{1}}v_{2} \nonumber \\
	    	& -v_{2}\partial_{x_{1}}(|\Psi |^{2}v_{1})-\frac{|\Psi |^{2}}{m_{2}} \partial_{x_{2}} (V+Q).
	\end{align}

	\noindent
	As earlier we substitute the sum of the classical and quantum potentials from (13) and then differentiate (12) over time, (18) over $x_{1}$ and (19) over $x_{2}$ and subtract from the first the sum of the two latter and we obtain finally

		\begin{align}
	  \partial_{t}^{2} |\Psi|^{2}= &\sum _{i=1}^{2}[\partial ^{2}_{ x_{i}}\left(|\Psi|^{2}v_{i}^{2}\right)-\partial _{ x_{i}}|\Psi|^{2}(\partial _{t}+v_{i}\partial _{x_{i}})v_{i}] \nonumber \\  
& +[\partial _{x_{1}}v_{1}\partial _{x_{2}}(|\Psi|^{2}v_{2})+v_{1}\partial _{x_{1}}\partial _{x_{2}}(|\Psi|^{2}v_{2})+(1\leftrightarrow 2)].
	\end{align}

	\noindent
	This is the two-body extension of the single particle wave-like equation of motion derived earlier. The last  term in the above expression is highly important due to the inclusion of spatial derivatives acting simultaneously at two distant points which indicates the non-separability of the bi-particle compound. It exhibits particular conceptual interest as the gradient operators act at two distant sites and cannot be ascribed to one or the other individual particle and the inter-particle expressions for the many-body case follow the same pattern. Even though such kinds of derivatives also appear in the Pauli-Jordan field commutation relations \cite{BohrRosenfeld} their physical significance is associated with limitations in the simultaneous measurability of field averages for space-time connected finite space-time regions. As it has been shown in particular by Bohr and Rosenfeld in their seminal paper, the disturbance caused by a scalar electric field induced by the test body is responsible for these spatial derivative terms. The startling difference is that in the present case the relevant points may be very well casually disconnected so no such influence is possible, which clearly undermines one of the basic premises of Bell's premises which is separability and shows there is considerable tension between quantum mechanics and the field theory description. It should be added furthermore that the ontological status of particle trajectories, which has such a prominent place in Bohm's interpretation, here is severely undermined since there is an immediate effect between distant non-crossing trajectories. Another point to consider is that the expression given in (20) depends inversely on the two interacting particle masses. When $m_{1}\neq m_{2}$ it is reasonable to assume that $S_{12}\neq S_{21}$ from which it follows that when we interchange indices (20) does not remain invariant so $|\Psi_{12}|^{2}\neq |\Psi_{21}|^{2}$. In a sense it is the non-invariance of the unobservable phase function that determines the asymmetry in the observable probability density. It is clear, and we will return later to this point, that this is not the case for identical particles with equal masses. It is evident that an N particle compound corresponds to a sum of $N(N-1)/2$ terms of that kind and when the mass of a particle is macroscopic its contribution will be negligible. What is clearly violated is the principle of spatio-temporal separability and the individuation of particles associated with the bi-particle which seems to constitute a single irreducible entity. Accordingly, the measure of non-classicality may be quantified by exactly that kind of inter-particle terms. Those terms that include gradients acting independently at two distant points express the classicality of the system, but those rest of them involving inter-site derivatives illustrate its unique non-classical character. Finally, we can explore the implications following from the two-particle equation of motion we obtained and its relation to the permutation invariance principle. It follows after expanding the derivatives in (20) that
	
\begin{equation}
\left(\partial ^{2} _{t}  -\hat{\Lambda}\right)\left(|\Psi _{12}|^{2}-|\Psi _{21}|^{2}\right)=0,
	\end{equation}

	\noindent
	where $\hat{\Lambda}$ a linear operator acting on the coincidence probability density which is invariant if we permute the two particle indices, so that $\hat{\Lambda}=\hat{\Lambda}_{12}=\hat{\Lambda}_{21}$. It follows directly then from (21) that $|\Psi _{12}|^{2}=|\Psi _{21}|^{2}$. We see then that Born's rule dictates a linear equation for the probability density and consequently ensures its invariance under particle permutations, so it is not possible to accept stochastic fluctuations that may generate deviations from the guiding velocity law and in addition the equilibrium condition stated earlier. Aside from this kind of argumentation, we could also maintain that a possible variance from this rule would not automatically ensure the validity of the permutation invariance \cite{French} of the coincidence probability density which is a conclusion not acceptable on physical grounds. Incidentally, this kind of reasoning also excludes the possibility of instantaneous signalling  \cite{Valentini2}. 
	
	\section{Concluding remarks}
	
	Without modifying Schr\"{o}dinger's equation we sought an equation of motion for the probability density with second order time derivatives that admits solutions which satisfy the condition $\rho=|\Psi|^{2}$. The desired wave equation contains some intriguing terms that signify non-classical inter-particle correlations and the interconnectedness of the quantum compound. It was suggested that a dual ontology is indicated by the presented formalism. On one hand the two sum terms in (20) correspond to the two distinct single particles while the inter-particle cross-terms signify the bi-particle presence and the unicity of the quantum aggregate. We have also proved that as a consequence of Born's rule  the probability density remains invariant under the permutation of the constituent particles of the quantal compound. This conclusion follows directly from the linearity of the operator $\Lambda$, so it is based solely on theoretical considerations and reveals a deep underlying relationship between the quantum equilibrium hypothesis \cite{Cushing2, Valentini2} and permutation invariance \cite{French}. In addition we have argued that a possible gradual transition to quantum equilibrium $\rho \rightarrow |\Psi|^{2}$, through stochastic or other relaxation processes \cite{Bohm} seems implausible.


\begin{thebibliography}{99}
	
		
		
		\bibitem{Dyson} F. Dyson, Why is Maxwell's Theory so hard to understand?, James Clerk Maxwell Foundation's Commemorative Booklet. Edinburgh (1999)  
		
		
		\bibitem{Cushing} J. T. Cushing, Philosophical Consequences of Quantum Theory: Reflections on Bell's Theorem,	J. T. Cushing and E. McMullin (eds.),
		University of Notre Dame Press;  (1989)
		
		\bibitem{Wyatt} P. R. Wyatt, \emph{Quantum Dynamics with Trajectories: Introduction to Quantum Hydrodynamics} (Springer, 2006)
		
		\bibitem{Bohm} D. Bohm and B. Hiley \emph{The Undivided Universe} (Routledge,1993)
		
		\bibitem{Holland} P. R. Holland, \emph{The Quantum Theory of Motion} (Cambridge University Press, 1995)
		
		\bibitem{Cushing2} J. T. Cushing: Quantum Mechanics: Historical Contingency and the Copenhagen Hegemony, University of Chicago Press, Chicago (1994)
		
		\bibitem{aeroacoustics} X. Sun and X. Wang, Fundamentals of Aeroacoustics with Applications to Aeropropulsion Systems,  Academic Press (2020)
		
		\bibitem{Ghirardi} 	E. Deotto  and G. C. Ghirardi, Bohmian Mechanics Revisited, \emph{Foundations of Physics}, \textbf{28}, 1 (1998)
		
	
		
		\bibitem{BohrRosenfeld} N. Bohr and L. Rosenfeld.: On the question of the measurability of electromagnetic field quantities. Reprinted in J. A. Wheeler and W. H. Zurek, (eds.): Quantum theory and measurement, p. 479, Princeton University Press, Princeton (1983)
		
		
		\bibitem{French} S. French and D. Krause: \emph{Identity in Physics: A Historical, Philosophical, and Formal Analysis}. Oxford University Press, Oxford (2006)
		
		\bibitem{Valentini2} A. Valentini: Signal-locality, uncertainty, and the sub-quantum H-theorem, II, \emph{Physics Letters A}, \textbf{158}, 1, (1991)
		
		
		
		
		
	
		
		
	\end{thebibliography}
\end{document}